\documentstyle[jkas]{article}

\beginpage{1}
\endpage{4}

\runningauthor {T. C. Hinse et al.} 
\runningtitle{Transiting planets by telescope defocusing} 

\month{Feb} \year{2015} \volume{} \issueno{}
\beginpage{1}\endpage{9}

\date{Received ... 2015; Accepted ... 2015}

\begin{document}

\title{Photometric defocus observations of transiting extrasolar planets} 

\author{T.~C.~Hinse$^{1,2}$, Wonyong Han$^{1}$, Jo-Na Yoon$^3$, \\
Chung-Uk Lee$^{1}$, Yong-Gi Kim$^{4}$, Chun-Hwey Kim$^{4}$} 

\address{$^1$ Korea Astronomy and Space Science Institute, Daejeon 305-348, Republic of Korea.\\
$^2$ Armagh Observatory, College Hill, BT61~9DG Armagh, United Kingdom.\\
$^3$ Chungbuk National University Observatory, Cheongju 361-763, Republic of Korea.\\
$^4$ Chungbuk National University, Cheongju 361-763, Republic of Korea.}
\offprints{Wonyong Han, whan@kasi.re.kr}

\abstract{We have carried out photometric follow-up observations of bright transiting extrasolar planets using the CbNUOJ 0.6m telescope. 
We have tested the possibility of obtaining high photometric precision by applying the telescope defocus technique allowing the use of 
several hundred seconds in exposure time for a single measurement. We demonstrate that this technique is capable of obtaining a 
root-mean-square scatter of order sub-millimagnitude over several hours for a V $\sim$ 10 host star typical for transiting planets detected 
from ground-based survey facilities. We compare our results with 
transit observations with the telescope operated in in-focus mode. High photometric precision is obtained due to the collection of a 
larger amount of photons resulting in a higher signal compared to other random and systematic noise sources. Accurate telescope tracking is likely to 
further contribute to lowering systematic noise by probing the same pixels on the CCD. Furthermore, a longer exposure time helps reducing 
the effect of scintillation noise which otherwise has a significant effect for small-aperture telescopes operated in in-focus mode. Finally 
we present the results of modelling four light-curves for which a root-mean-square scatter of 0.70 to 2.3 milli-magnitudes have been achieved.}


\maketitle

\section{Introduction}

The first discovery of an extrasolar planet in 1995 \citep{MayorQueloz1995} resulted in the opening of a completely new astronomical research area. However, detailed information of the planet obtained via radial velocity measurements is limited as only the minimum mass (among other parameters) could 
be inferred due to an unknown orbital inclination. The class of transiting extrasolar planets has changed this ambiguity. The discovery of a transiting extrasolar planet (TEP) \citep{Charbonneau2000,Henry2000} allows the determination of physical properties of the planet (mass, radius among others) and its host star (limb-darkening, effective temperature among others) 
when combined with spectroscopic observations. In principle, obtaining precise measurements of these properties will distinguish a given TEP to be a rocky terrestrial-size or gaseous Jupiter-size planet by inferring its absolute size and mass. Therefore accurate properties will help to constrain planet formation theories. One possible technique to obtain high-precision photometric light-curve is the use of telescope defocus technique and has been successfully applied in various studies \cite{SouthworthWASP52009Paper}. This research presents results from our attempt to obtain precise light-curves of TEPs using a 0.6m telescope. In particular we present results for which the telescope was operated in-focus as well as out-of-focus. In section 2 we outline the background of the defocus technique. In section 3 we present our observations of seven TEPs observed since early 2012. A description of data reduction and light-curve modelling is given in section 4 and 5. Our results and analysis are given in section 6 and 7 followed by a conclusion in section 8.

\section{Defocus photometry}

Traditional astronomical photometry aims to measure the brightness of a star with the telescope kept in focus resulting in a well-defined point-spread function (PSF). This is usually desired when structural information is needed (e.g. star cluster observations to resolve individual stars) or when the observed star field is crowded to avoid confusion due to overlapping PSFs. However, in certain situations the operation of the telescope in a defocus mode opens up for the possibility to achieve an increased signal-to-noise (S/N) ratio resulting in a higher photometric precision. Nevertheless the same noise processes are in operation whether the telescope is in-focus or de-focused. To obtain high-S/N measurements over an extended time period both instrumental and environmental effects are important to consider. 

In general several types of noise sources contributes to the overall error budget of a single measurement. The first class is random noise (also known as statistical, white or time-uncorrelated noise) and often referred to Gaussian noise associated with Poisson counting statistics of photons. CCD read-out noise, CCD dark-current, shot-noise of the background as well as the target star are all referred to random noise. In a special case atmospheric scintillation noise is also of random nature, but depending on other factors such as telescope aperture (see below for more details). To decrease white noise it requires an increase in the counting statistic of photons. The second class of noise source is of systematic nature and harder to identify and/or quantify. Examples of systematics would be selecting aperture sizes for photometry or flat-fielding errors involving the positioning of the target star on the same pixels. Controlling the latter is ultimately linked with telescope pointing stability and involves how well the telescope is tracking over an extended time period. Also the choice of comparison stars for differential photometry is a potential source of systematic noise. Furthermore, the telescope optical alignment (astigmatism) and general telescope optics also are potential sources for introducing systematic noise. 

\begin{figure}[!t]
\centering
\epsfxsize=8cm 
\epsfbox{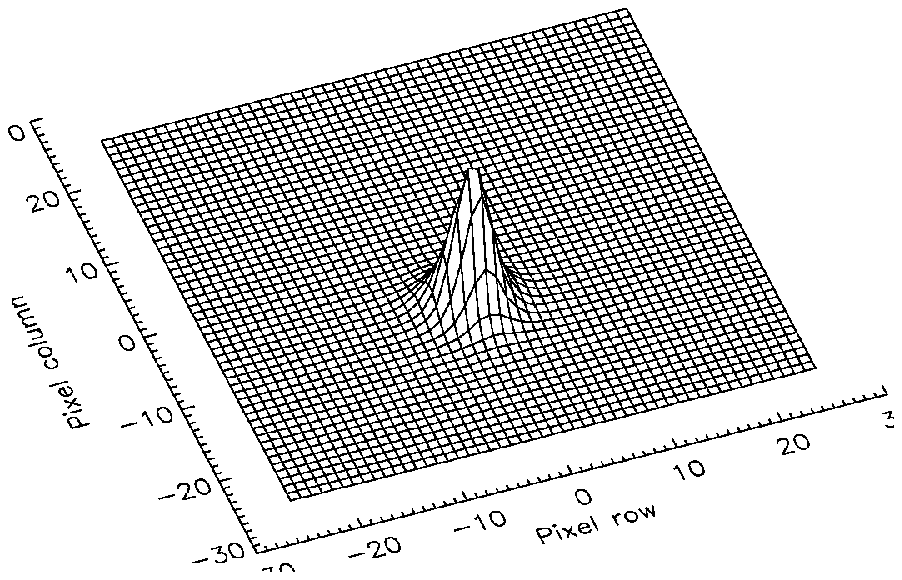} 
\epsfxsize=8cm
\epsfbox{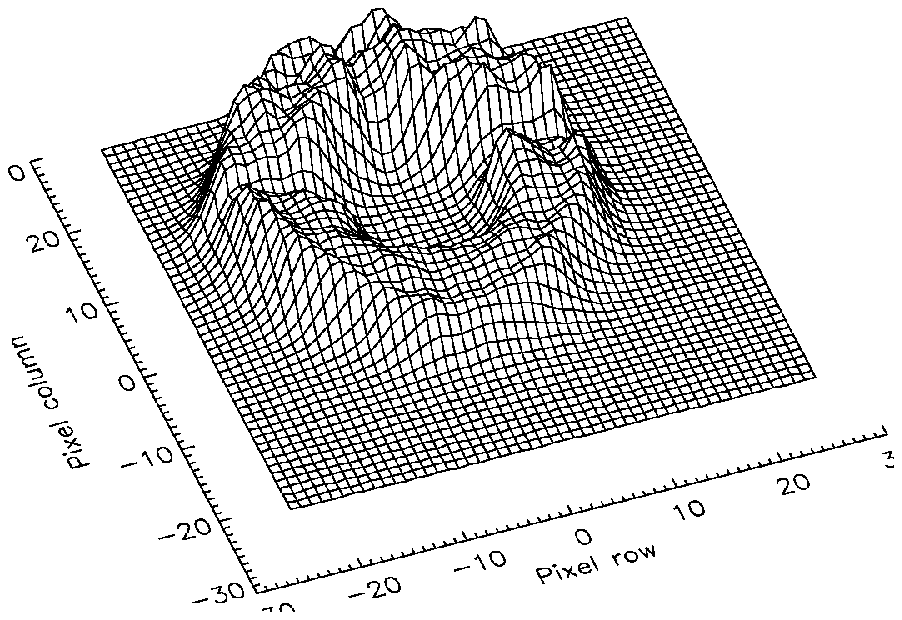}
\caption{Surface plots of two PSFs. Upper panel: PSF of XO3 (2012/02/11) with the telescope well focused and an exposure time of 12s. Lower panel: PSF of HATP22 (2014/01/13) with the telescope defocused heavily allowing for a much longer exposure time of 195s. Both stars are of same brightness (see Table \ref{HostStarDetails}).}
\label{PSF}
\end{figure}

\begin{figure}[!t]
\centering
\epsfxsize=8.5cm 
\epsfbox{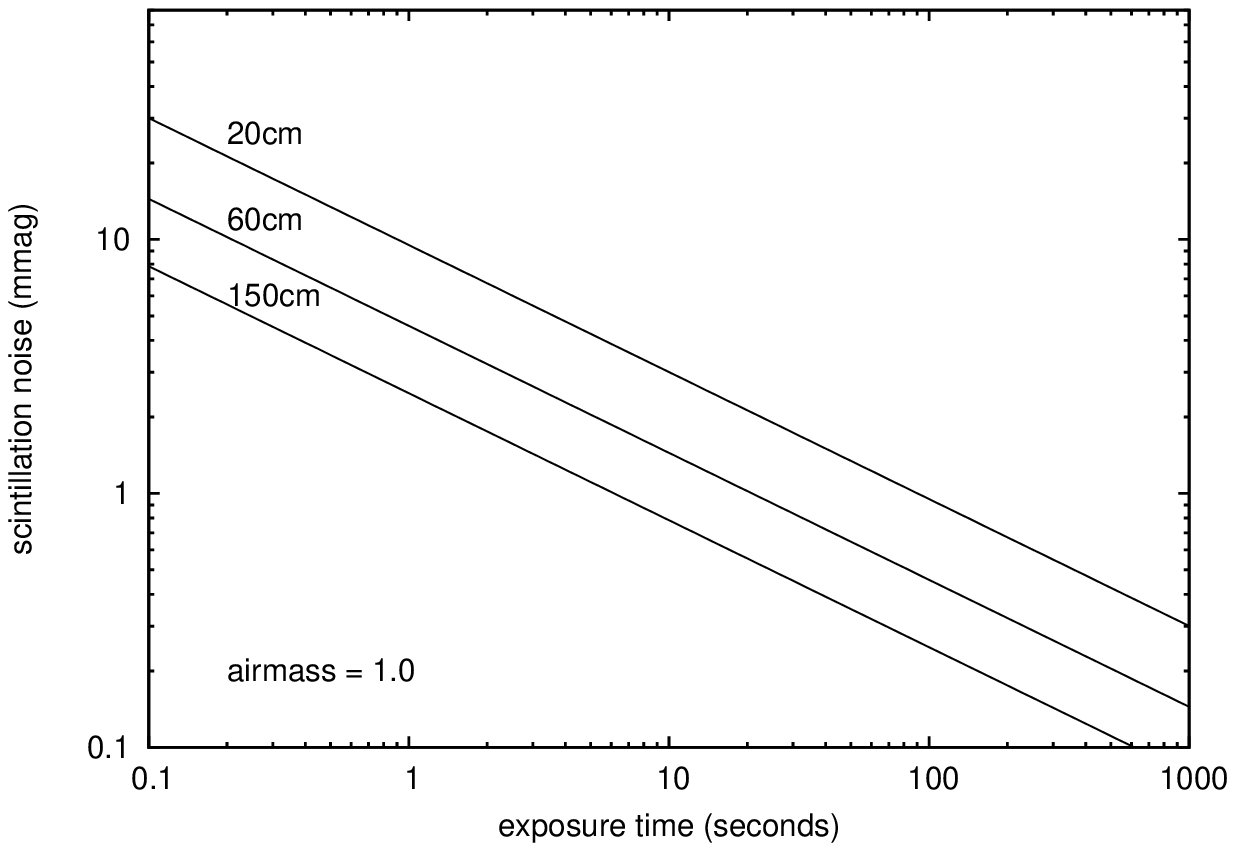} 
\epsfxsize=8.5cm
\epsfbox{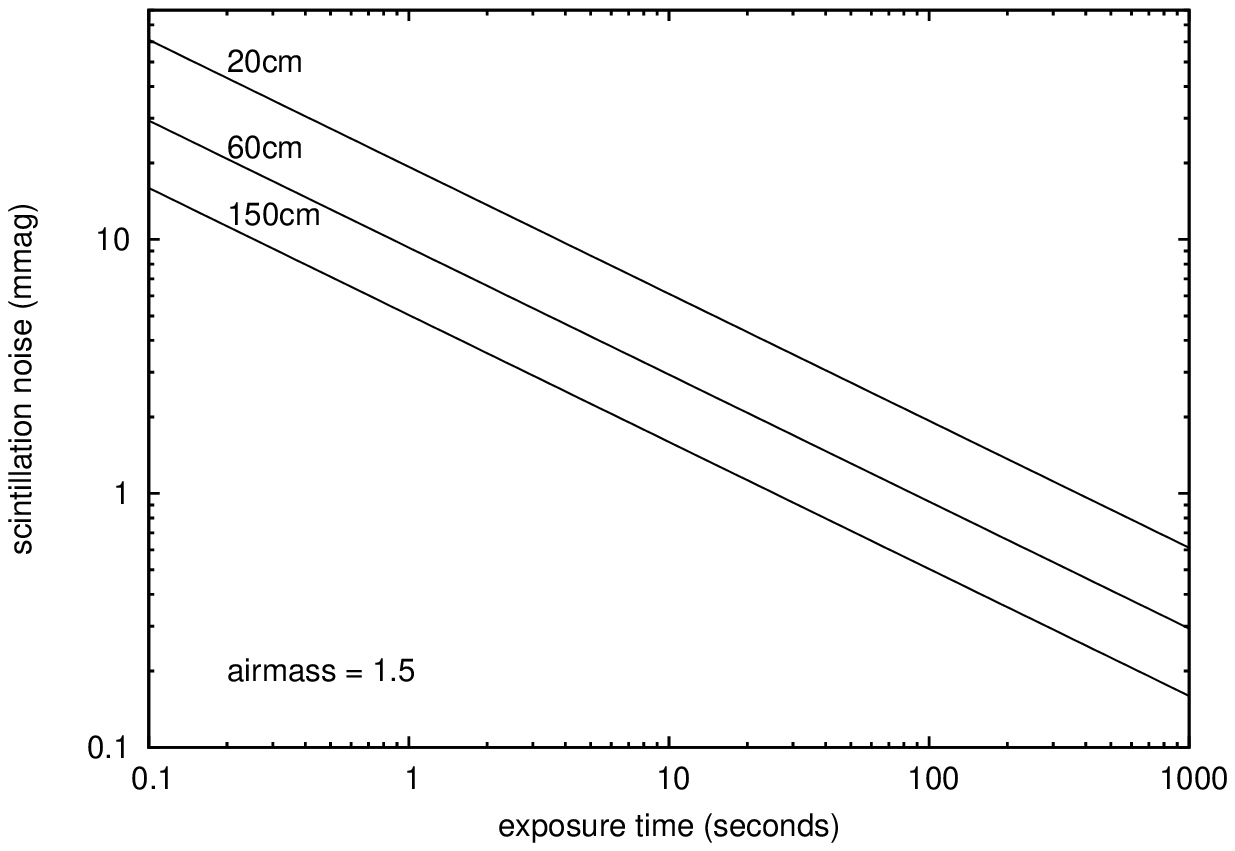}
\caption{Expected photometric error due to scintillation noise as a function of exposure time and mirror aperture valid for a telescope at 80m above sealevel. \emph{Top panel}: Considering an observation through air-mass 1.0. \emph{Lower panel}: Considering air-mass of 1.5. See text for more details.}
\label{ScintillationPlots}
\end{figure}

Other noise sources include non-random noise (also known as red, $1/f$ or time-correlated noise) often (but not exclusively) associated with atmospheric extinction encountered at the beginning and/or end of a transit observation when the target star is observed through high air-mass. The final error source is related to the intrinsic variability of the star itself such as star spots, flares and pulsations. All of these noise sources have the effect of deteriorating the photometric time-series measurements in one or the other way making it difficult to obtain a photometric precision of less than a milli-magnitude.

One key aspect of limiting the S/N of a photometric measurement is the dynamic range of a CCD detector dictating the peak flux and thereby setting a natural upper limit on the exposure time for a telescope operated in a in-focus mode. CCD detectors with a higher dynamic range allow for a longer exposure time thereby circumventing the danger of reaching the saturation limit. Ideally, in order to obtain a high signal-to-noise (S/N) ratio plenty of photons are needed (otherwise achieved by using a large mirror aperture). The larger the photon counts the lower the Poisson noise. One practical way to allow for a longer exposure time without saturating the CCD is to spread the light over many pixels by defocusing the telescope. As a result the stellar PSF of the target (and all other stars in the same field) broadens and decreases the peak value of an exposed pixel. 

In Fig.~\ref{PSF} we show an example of two stars of fairly similar brightness, but defocused for the case of HATP22. Simply, in a defocus mode, the number of photons are distributed over a larger area of the CCD allowing a prolonged exposure and hence a stronger signal to build up and stand out from the baseline of combined noise sources. 

However, a natural upper limit of exposure time $(t_{exp})$ is determined by the single frame read-out time $(t_{ro})$ and the transit duration $(t_{dur})$ as follows

\begin{equation}
T_{dur} = N \times (t_{exp}+t_{ro}) 
\end{equation}
\noindent
where $N$ is the number of frames recorded during the transit. To characterise a planetary transit from a detailed modeling process the light-curve sampling has to be sufficiently high in order to ensure enough measurements. As a rule of thumb we aim to obtain $N = 30$ to $40$ measurements during a given transit event resulting in 4 to 6 measurements of the ingress and egress phase.

Several advantages and disadvantages of the defocus technique contribute to either decrease or increase the S/N ratio. The variable atmospheric seeing effect and local temperature changes are less important for observations with a broad PSF. However, a high atmospheric transparency is still desireable. Distributing the photons over many pixels also has the effect of decreasing the photometric noise due to intrinsic CCD pixel-to-pixel response variations. The total flatfield noise contribution within a larger signal aperture (contribution of each pixel added in quadrature) is lower when compared to a similar in-focus measurement. In essence, the flatfield error of the final photometric measurement should average out using a larger number of independent pixels due to its random nature. One technical possibility along which one can increase the number of pixel is to operate the CCD in one-by-one binning. As the maximum number of pixels are requested the read-out time in single-pixel binning mode is maximised. Forming super pixels via two-by-two (or higher) binning would decrease the read-out time. However, a disadvantage of defocus photometry is the contribution of CCD read-out noise. The more pixels are exposed the larger the total read-out noise on a given measurement. In addition, the contribution of background noise in the broadened PSF also will decrease the photometric S/N. 

A positive effect of a prolonged exposure for a small-aperture telescope is the beating of scintillation noise. Scintillation is caused due to the refraction of starlight (twinkle) by turbulent cells in the atmosphere. The effect introduces a time-varying fluctuation in the star's brightness. Its effect is largest for telescopes with small mirror diameters and short exposure times. Therefore, scintillation will set a lower-limit of the photometric precision obtained for a small-aperture telescope. The first to model the effect of scintillation for bright stars is \cite{Young1967}. Using the expression given in \cite{Hartman2005} we have examined the dependence of scintillation noise on telescope aperture, air-mass, telescope height and exposure time. In Fig.~\ref{ScintillationPlots} we plot the scintillation noise (in milli-magnitude) as a function of exposure time for three different telescope apertures located at 80m above sealevel. The top panel shows the relationship for 
air-mass=1.0 and the lower panel for air-mass=1.5. In general, the scintillation noise decreases for either increasing telescope aperture or for increasing exposure time at a given air-mass. Considering a 60cm telescope, at unit air-mass and integrating for 250 seconds the scatter per datapoint expected from scintillation is below 0.3 mmag. At air-mass 1.5 this limit increases to 0.6 mmag - more than sufficient to record transits depths of 1 - 2\% with good precision. However, scintillation imposes a lower limit which is achievable for only bright stars.

\begin{table*}[t]
\begin{center}
\centering
\caption{Observation log of seven TEPs. The transit of HATP09 was the first transit to be recorded using the new ST-16083 camera.}
\begin{tabular}{ccccccccc} 
\hline
Target & Date & Start time & End time & Exposure time & Filter & \# Obs. & Air mass & Camera \\
\hline\hline
XO5 & 2012/02/10 & UT 10:09 & UT 18:25 & 75s - 105s & R & 288 & 1.00-1.56 & ST-8\\
XO3 & 2012/02/11 & UT 09:58 & UT 15:06 & 10s - 12s & R & 1084 & 1.08-1.86 & ST-8\\
XO4 & 2012/02/17 & UT 10:55 & UT 18:05 & 27s - 35s & R & 568 & 1.12-1.63 & ST-8\\
HATP25 & 2012/11/15 & UT 10:49 & UT 18:50 & 170s - 190s & R & 149 & 1.02-1.83 & ST-8 \\
HATP09 & 2012/12/23 & UT 12:33 & UT 20:22 & 230s (fixed) & R & 109 & 1.00-1.62 & ST-16083\\
HATP22 & 2014/01/13 & UT 14:41 & UT 21:23 & 195s (fixed) & R & 114 & 1.03-1.35 & ST-16083\\
HATP12 & 2014/03/09 & UT 13:42 & UT 18:39 & 260s (fixed) & R & 65 & 1.00-1.66 & ST-16083 \\ 
\hline 
\end{tabular}
\label{Table1}
\end{center}
\end{table*}

\begin{table*}[t]
\begin{center}
\centering
\caption{Details of seven host stars and their known TEPs. Data was obtained from http://simbad.u-strasbg.fr/simbad/sim-fid and http://exoplanet.eu.}
\begin{tabular}{cccccc} 
\hline
Target & RA (J2000) & Dec (J2000) & V (mag) & Depth (mag) & Duration (min) \\
\hline\hline
XO5    & 07 46 51.95 & +39 05 40.5 & 12.1  & 0.014  & 193 \\
XO3    & 04 21 52.71 & +57 49 01.9 & 9.9   & 0.0048 & 173 \\
XO4    & 07 21 33.17 & +58 16 05.0 & 10.8  & 0.0108 & 264 \\
HATP25 & 07 21 33.17 & +58 16 05.0 & 13.2  & 0.0204 & 169 \\
HATP09 & 07 20 40.48 & +37 08 26.5 & 12.3 & 0.0126 & 206 \\
HATP22 & 10 22 43.59 & +50 07 42.0 & 9.8   & 0.0119 & 172 \\
HATP12 & 13 57 33.48 & +43 29 36.7 & 12.8  & 0.0204 & 140 \\
\hline
\end{tabular}
\label{HostStarDetails}
\end{center}
\end{table*}

\section{Observations}

All observations\footnote{photometric data can be obtained in ASCII form from the corresponding author} presented in this work were carried out using the 0.6m telescope (hereafter CbNUOJ) located in Jincheon at an altitude of 87m above sealevel. The telescope was installed by Korea Astronomy and Space Science Institute (KASI) and is operated by Chungbuk National University Observatory, Republic of Korea. The telescope optics follows the Richey-Chr{\'e}tien design attached to an equatorial mount. Telescope tracking is performed via a servo motor control. More details of this telescope was described by \cite{Kim2014}.

During the observing period the telescope was equip\-ped with two different CCD imagers located at the Cassegrain f/2.92 prime focus and installed by the Korea Astronomy and Space Science Institute (KASI). Initially we used the 
1530 x 1020 pixel SBIG ST-8XE\footnote{http://archive.sbig.com/sbwhtmls/st8.htm} which was later replaced (currently installed) by the more advanced 4096 x 4096 pixel SBIG STX-16083\footnote{https://www.sbig.com/products/cameras/stx/stx-16803} CCD camera. The former camera had a field of view of $27^{'} \times 18^{'}$ and the current camera provides a field of view of $72^{'} \times 72^{'}$. Both cameras had a pixel scale of 1.05 arcsec/pixel. Single Johnson/Cousins UBVRI filters\footnote{www.astrodon.com} are available via an electronic filter wheel system. For most of our observations we used the Cousins R-filter. In some observations we experimented with clear filters. The large field of view in the current instrumental setup is ideal for differential photometry when carrying out ensemble photometry analysis \citep{EverettHowell2001}. Both cameras are cooled electronically. 

Unbinned CCD read-out time for the ST-8XE was about 10 seconds while for the STX-16083 camera it was around 16 seconds. The two relatively short read-out times are favourable for high-cadence transit photometry and allowed us to stay on target for a larger fraction of time for photon collection. Another benefit of short read-out times is a smaller contribution to the noise budget due to read-out noise imposed on a single frame as the high photon counts will dominate for defocus measurements. In the beginning (2012 season using the ST-8 camera) we followed a strategy where we changed the exposure time during a given night according to weather conditions. Later we abandoned this strategy and kept the exposure time fixed for consistency of the recorded data. Usually to our experience even thin clouds will have a temporal dramatic effect on the sampled light-curve when trying to aim for a mean $\sim 1$ mmag photometric precision over 3 to 4 hours. Changing exposure times by a few percent will not have a significant impact on the signal-to-noise ratio and our goal was to ensure a well-sampled light-curve. All observations were carried out with the cameras operated in unbinned mode providing a maximum number of pixels favourable for defocus photometry. Prior to the defocus observations we aimed to obtain 2-3 in-focus images in order to identify any nearby companions. The average seeing at CbNUOJ is around 2 arcsec.

Calibration frames (bias and skyflats) were obtained for each night using appropriate filter. We carried out tests for which the science images were either calibrated or kept in their original form. We will report about the results from those experiment in the next section. An observation log is given in Table \ref{Table1}. Details of the observed host stars are given in Table \ref{HostStarDetails}.

\section{Data reduction}

We used the \texttt{DEFOT}\footnote{http://www.astro.keele.ac.uk/$\sim$jkt} data reduction pipeline outlined in \citet{SouthworthWASP52009Paper}. The software is written in \texttt{IDL} and makes use of the NASA IDL astronomy library package \texttt{astrolib}. The method of obtaining photometric measurements is following the algorithm described in the \texttt{DAOPHOT} photometry package. Aperture photometry is done using the \texttt{astrolib/aper} routine. UTC (Coordinated Universal Time) time stamps in the format YYYY-MM-DD/HH-MM-SS are obtained from each image FITS header. Actual times for a single photometric measurement were taken to be the mid-exposure time. To convert to Barycentric Julian Date (BJD) in the Barycentric Dynamical Time (TDB) standard we used the \texttt{utc2bjd.pro} conversion routine\footnote{http://astroutils.astronomy.ohio-state.edu/time/} \cite{Eastman2010}. BJD time stamps expressed in the TDB standard will be particularly useful in when searching for transit timing variations (TTV).

Apertures for the inner and background annulus were fixed (but their optimum values determined from experimentation) in all frames for each data set. We noticed a significant frame-to-frame shift due to tracking imperfections of the telescope. To ensure that the apertures are following the selected stars (target and comparison) we made use of the image registration cross-correlation algorithm available in the \texttt{DEFOT} package. For each program image a summed pixel row and pixel column intensity profile is calculated and compared with the same profile of a reference frame. A least-squares high-order polynomium fit is then performed to calculate any shifts and a given frame is then corrected accordingly in order to match the reference frame. The downside of the package is that only a single aperture for the three annuli is applied to all stars. In practice this limits the selection of comparison stars similar in brightness (and hence angular extend on the CCD) with the target. From experimentation we used apertures that yielded the smallest root-mean-square (RMS) scatter around a best-fit model (see next section). Table \ref{Apertures} gives an overview of the various apertures used for each target.

Comparison stars were checked for internal variability and none were found for the selected stars. The \texttt{DEFOT} package allows to carry out ensemble differential photometry using weighted flux summation. This procedure minimises the overall Poisson noise while also avoids distortion in the transit light curve. More details of this procedure can be found in \cite{SouthworthWASP52009Paper} and \cite{EverettHowell2001}. 

Long-term trends over the observing window were removed using a polynomium fit to out-of-transit data. The resulting light curve was then normalized to zero differential magnitude. In all normalisations we applied a first-order 
polynomium fit. Higher-order terms would have the potential of introducing artifical distortions to the transit light curve.

In some cases we noticed obvious data outliers most likely due to passing thin clouds. These were removed prior to our photometric analysis. To find stable comparison stars we selected all suitable stars and carried out a first selection by eye. Then a subset of stable stars (4-15) were included in the final photometric analysis. To find optimum apertures we varied the three radii in a systematic way. For each experiment we found a best-fit solution with associated RMS scatter of the residuals. Radii resulting in a least scatter were chosen. These experiments were done using raw science frames. Using the most optimal apertures we also calibrated (bias subtraction and flatfielding) the science frames to determine any difference in RMS. In some cases we found that a calibration resulted in a slight improvement in the photometry in other cases 
we found no improvement.

\begin{table}[t]
\begin{center}
\centering
\caption{Final aperture radii used for photometric measurements. $R_1$ is the inner aperture. $R_{2},R_{3}$ are the apertures for the background measurement. All numbers are in pixels. The apertures were determined in order to minimise the RMS scatter of the data.}
\begin{tabular}{cccc} 
\hline
Target & $R_1$ & $R_2$ & $R_3$ \\
\hline\hline
HATP25 & 9 & 13 & 20 \\
HATP09 & 12 & 40 & 45 \\
HATP22 & 25 & 35 & 45 \\ 
HATP12 & 10 & 30 & 45 \\
\hline
\end{tabular}
\label{Apertures}
\end{center}
\end{table}

\begin{figure}[!t]
\centering
\epsfxsize=8.0cm 
\epsfbox{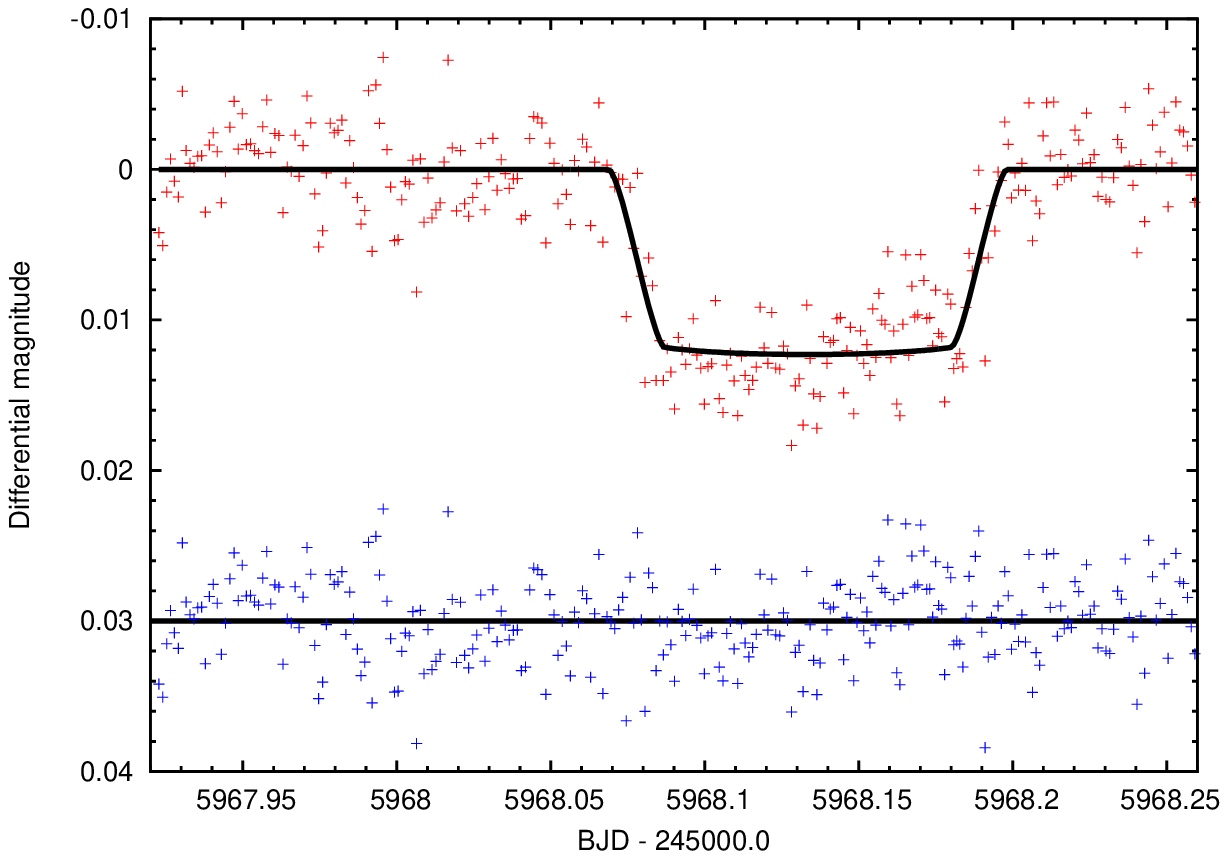} 
\epsfxsize=8.0cm 
\epsfbox{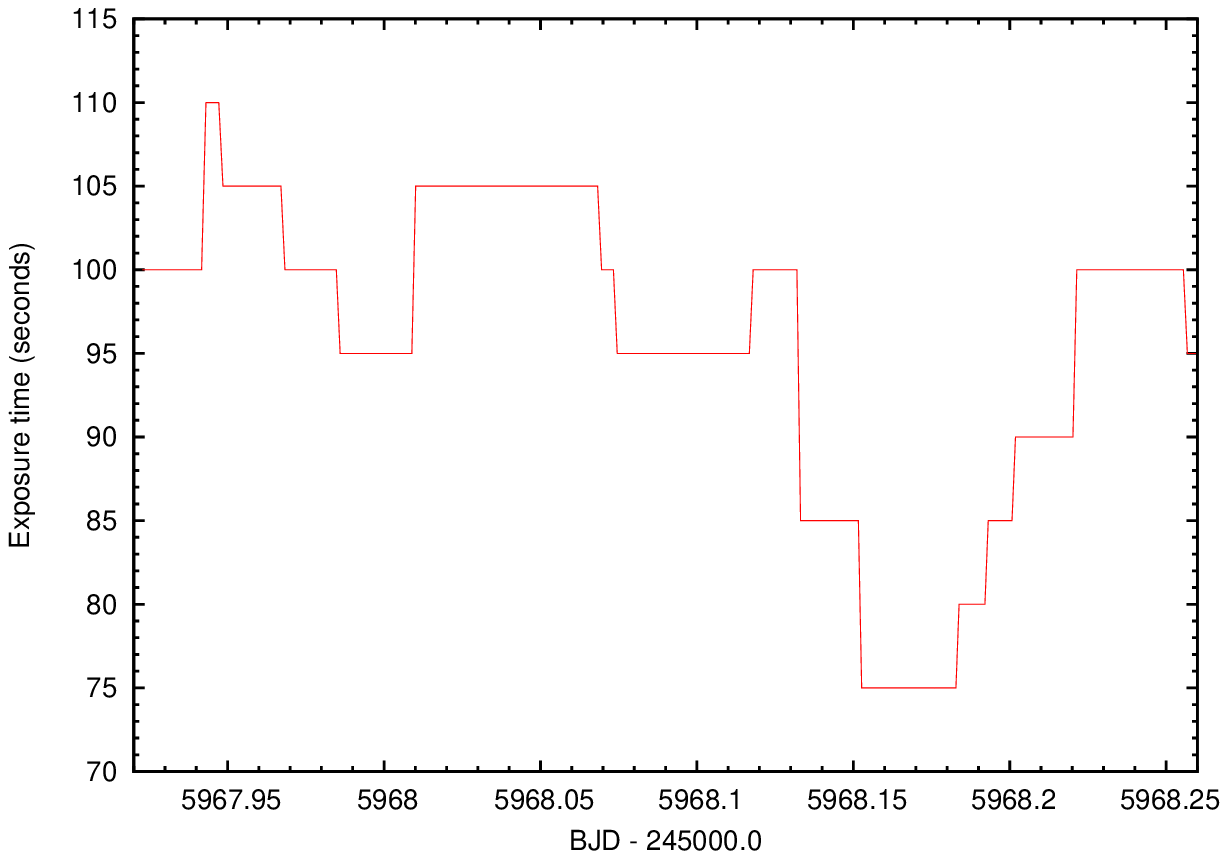} 
\caption{Light-curve of XO5 as observed on 2012/02/10 with the telescope in focus. Upper panel: The RMS scatter around the best-fit model is 2.98 mmag. The large scatter does not allow for a detailed limb-darkening treatment. Ingress and egress phases are not characterised well. The mid-transit time is at BJD
$2,455,968.13350 \pm 0.00061$. Lower panel: Various choices of exposure times during the observing period. Some correlations between short exposures and large scatter is visible.}
\label{XO5_LC}
\end{figure}

\begin{figure}[!t]
\centering
\epsfxsize=8.0cm 
\epsfbox{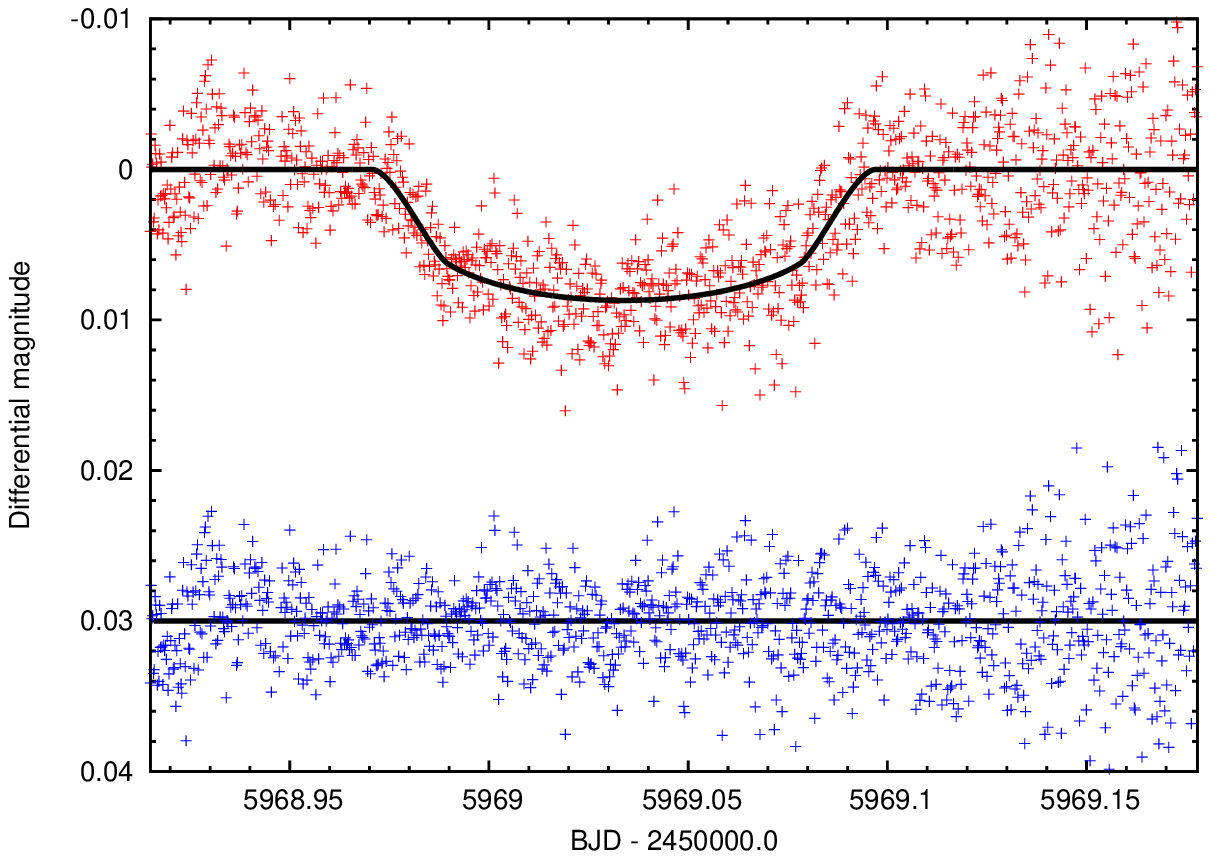} 
\epsfxsize=8.0cm 
\epsfbox{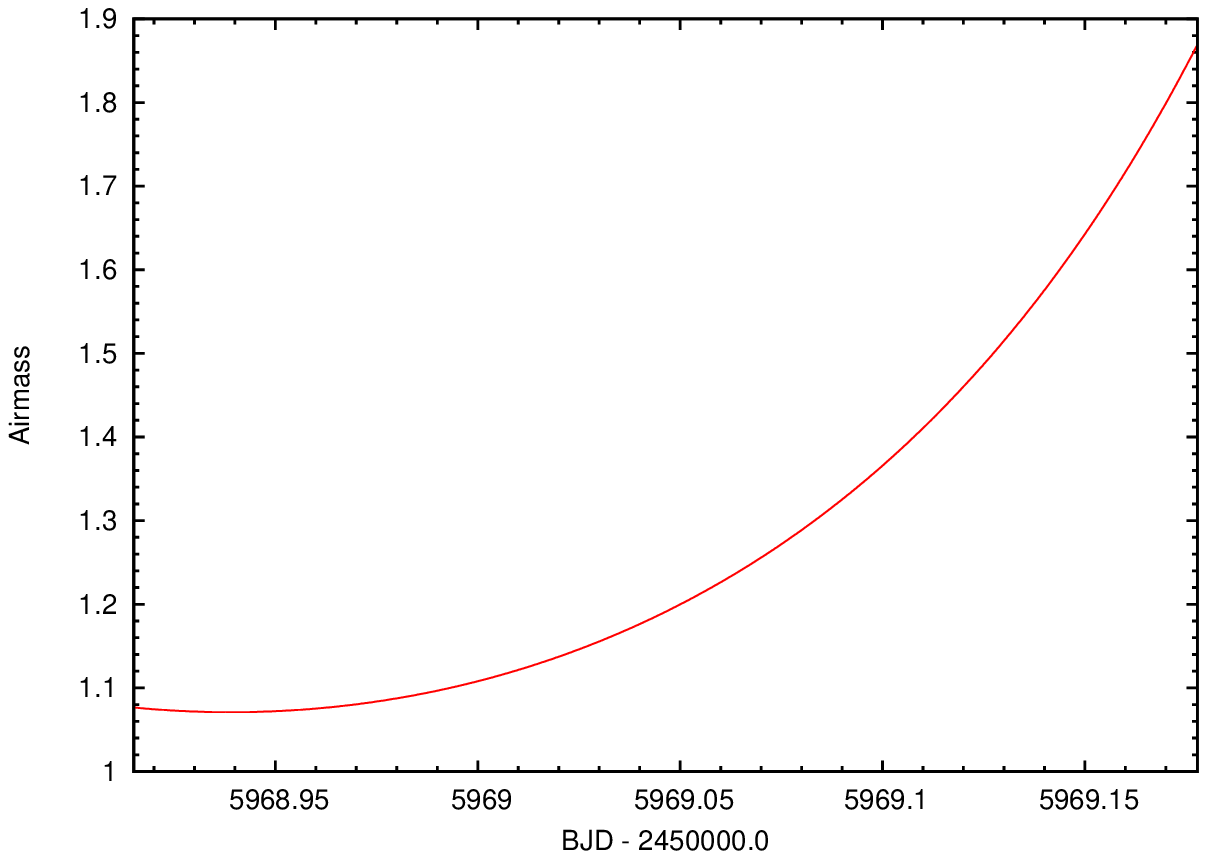} 
\caption{Light-curve of XO3 as observed on 2012/02/11 with the telescope in focus. Upper panel: The RMS scatter around the best-fit model is 3.18 mmag. The mid-transit time was determined as BJD $2,455,969.03386 \pm 0.00024$. Lower panel: Air-mass versus time. Due to high air-mass at the end of observing night the photometric scatter increases.}
\label{XO3_LC}
\end{figure}

\begin{figure}[!t]
\centering
\epsfxsize=8cm 
\epsfbox{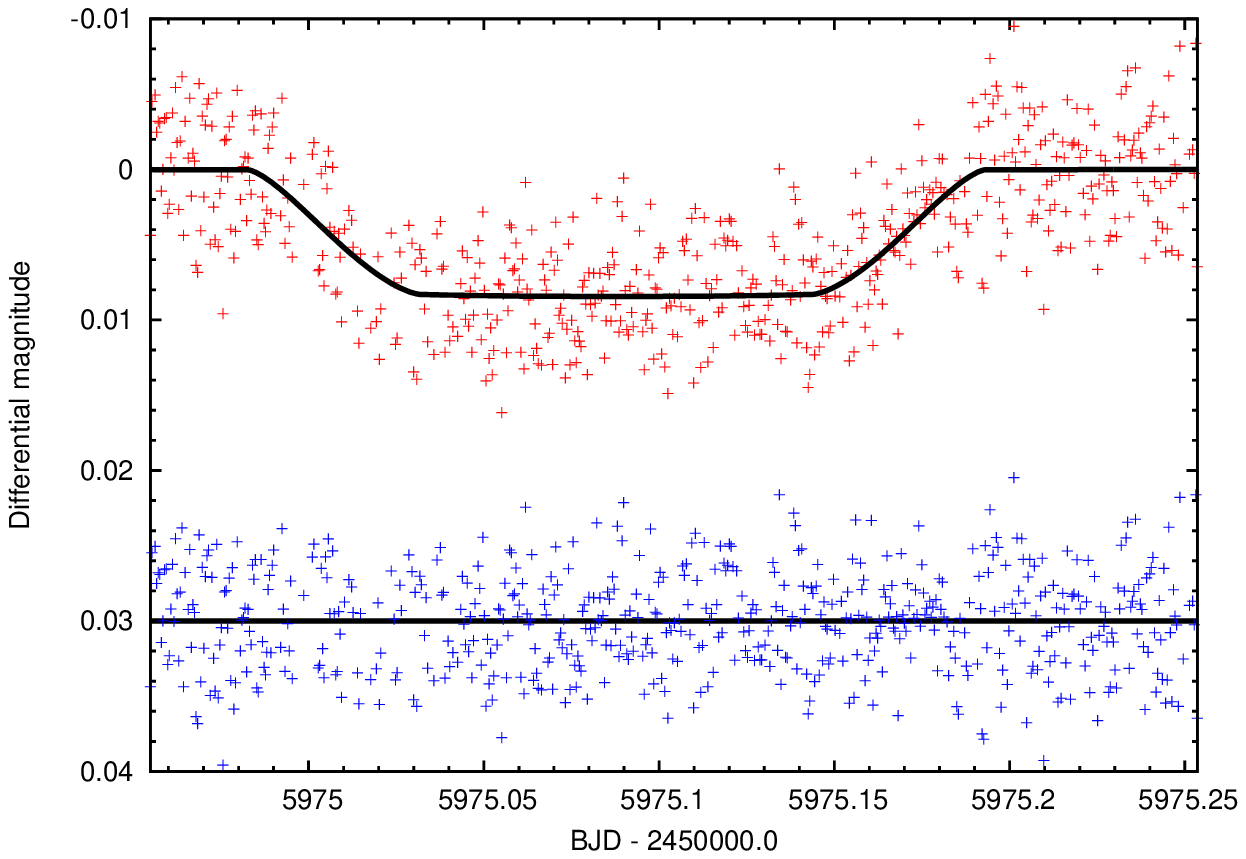} 
\epsfxsize=8cm 
\epsfbox{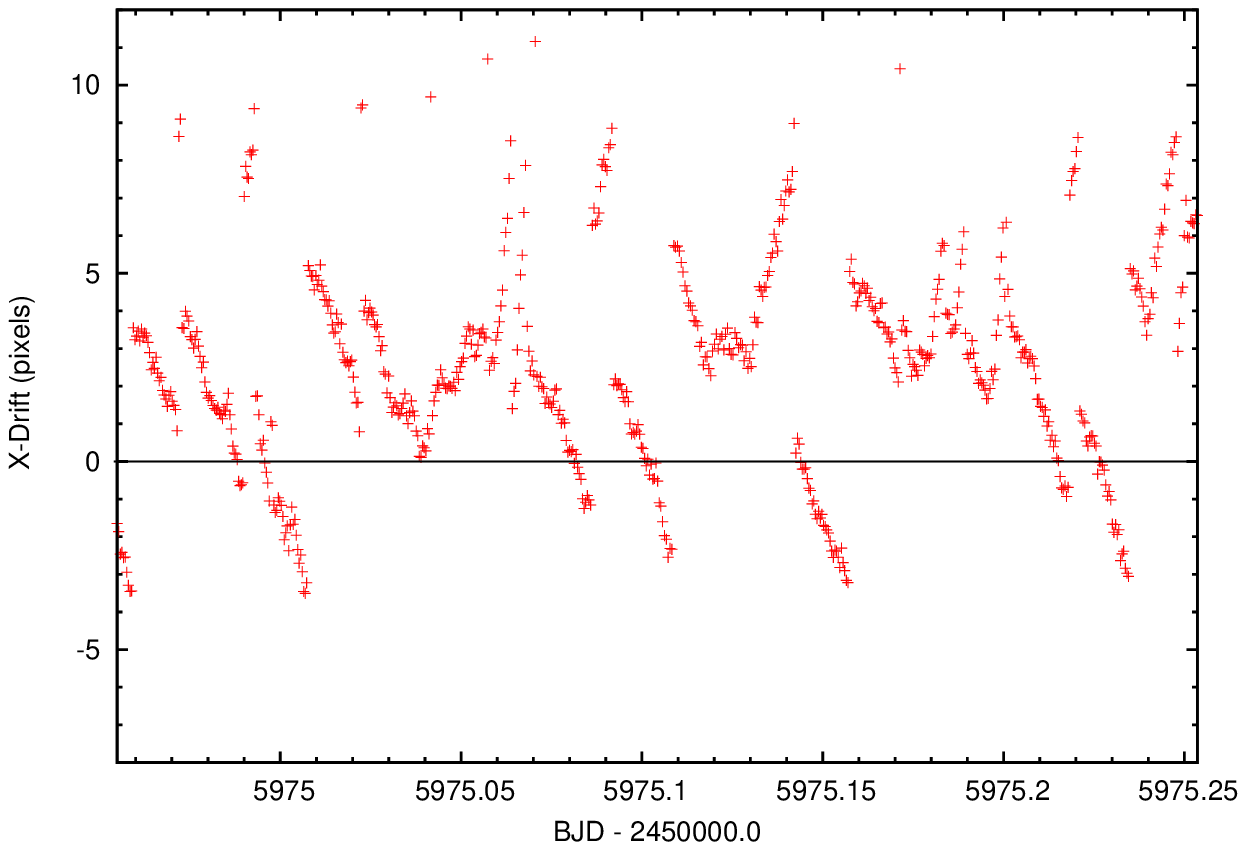} 
\epsfxsize=8cm 
\epsfbox{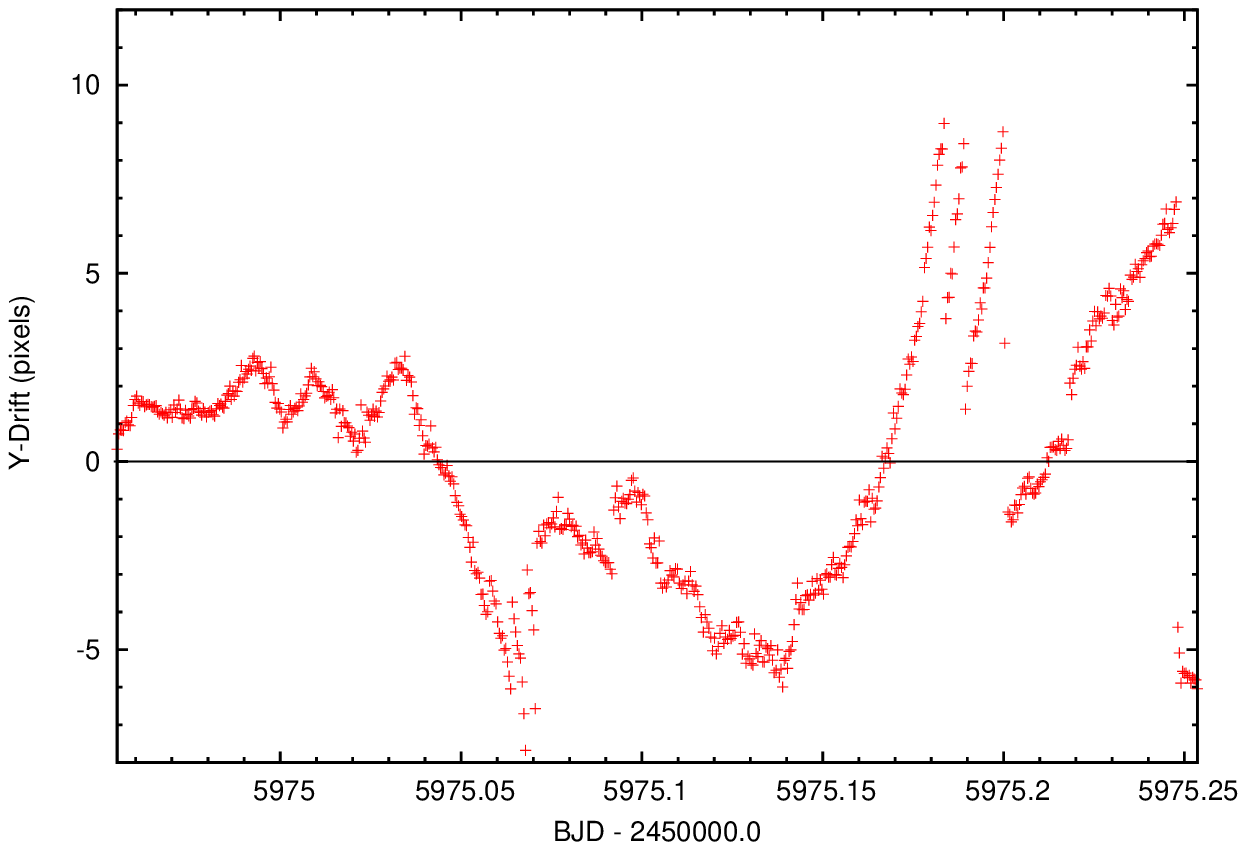} 
\caption{Light-curve of XO4 as observed on 2012/02/17 with the telescope in focus. Upper panel: The RMS scatter around the best-fit model is 3.34 mmag. The mid-transit time was determined as BJD $2,455,975.088 \pm 0.002$. Middel and Lower panels: Telescopic X and Y drift of program image relative to reference image.}
\label{XO4_LC}
\end{figure}

\begin{figure}[!t]
\centering
\epsfxsize=8.5cm 
\epsfbox{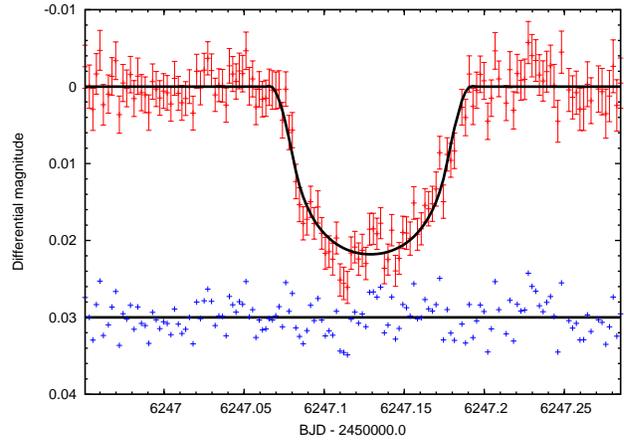} 
\caption{Light-curve of HATP25 as observed on 2012/11/15 with the telescope de-focused to allow for longer exposures. The RMS scatter around the best-fit model is 2.30 mmag with $\chi^2_{r}=1.01$. No calibration frames were applied.}
\label{HATP25_LC}
\end{figure}

\begin{figure}[!t]
\centering
\epsfxsize=8.5cm 
\epsfbox{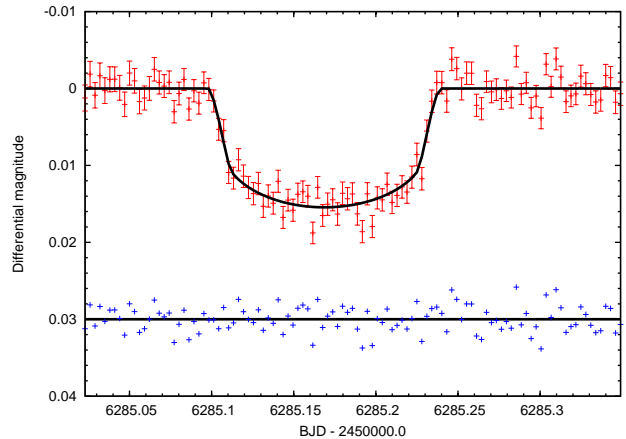} 
\caption{Light-curve of HATP09 as observed on 2012/12/23 with the telescope de-focused to allow for longer exposures. The RMS scatter around the best-fit model is 1.68 mmag with $\chi^2_{r}=1.42$. A linear limb-darkening law was determined to model the data best. No calibration were performed.}
\label{HATP09_LC}
\end{figure}

\begin{figure}[!t]
\centering
\epsfxsize=8.5cm 
\epsfbox{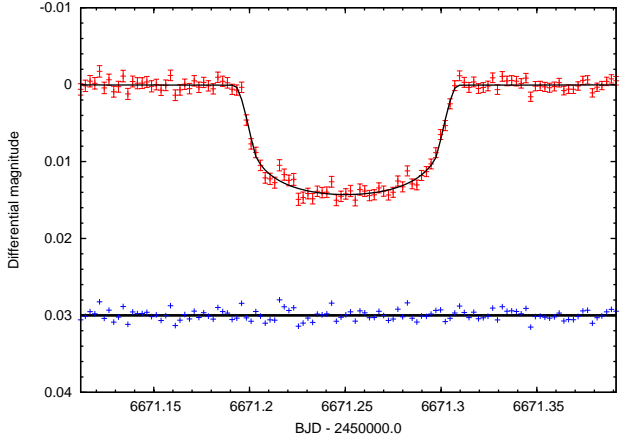} 
\caption{Light-curve of HATP22 as observed on 2014/01/13 with the telescope heavily de-focused to allow for longer exposures. The RMS scatter around the best-fit model is 0.70 mmag with $\chi^2_{r}=1.21$. Bias subtractions and flatfielding calibration were performed.}
\label{HATP22_LC}
\end{figure}

\begin{figure}[!t]
\centering
\epsfxsize=8.5cm 
\epsfbox{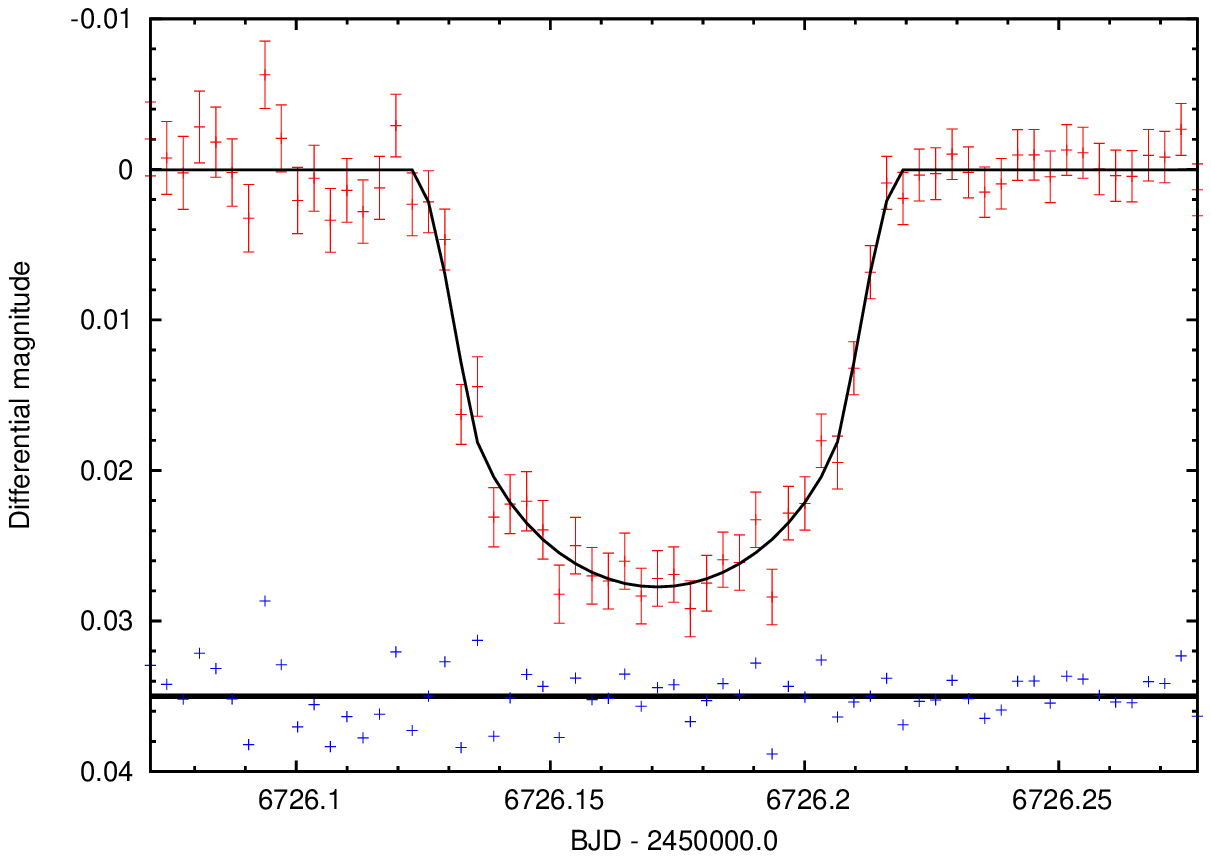} 
\caption{Light-curve of HATP12 as observed on 2014/03/09 with the telescope slightly de-focused to allow for longer exposures. The RMS scatter around the best-fit model is 1.85 mmag with $\chi^2_{r}=0.91$. No bias subtractions and flatfielding calibration were performed.}
\label{HATP12_LC}
\end{figure}

\section{Light-curve modelling}

Light-curves of all transiting planets were modelled using the \texttt{JKTEBOP\_V25}\footnote{http://www.astro.keele.ac.uk/$\sim$jkt} least-squares minimisation code \cite{Southworth2008,SouthworthWASP52009Paper}. The code models the light curve of detached eclipsing binary stars using biaxial spheroids. Fundamental parameters (star and planet are referred to with subscripts $A$ and $b$, respectively) describing the light-curve of a transiting planets are the orbital period $P$, the time of minimum light $T_0$ (or ephemeris epoch when modelling multiple light-curves simulteanously), the fractional radii of the star ($r_{A}=R_{A}/a$ where $a$ is the orbital semi-major axis) and planet ($r_{b}=R_{b}/a$), the orbital inclination ($i_{b}$) and limb-darkening coeffiecients. In practice, the parameters $r_A$ and $r_b$ are parameterised as their sum $(r_{A}+r_{b})$ and ratio $(k=r_{b}/r_{A})$. 

In all modelling work we kept the radii sum, radii ratio, orbital inclination, the time of minimum light and the scale factor (normalised flux level) freely adjustable. Since this work presents transit data of a single event for a given planet we did not fit for the orbital period and kept it fixed. Stellar limb-darkening is implemented in \texttt{JKTEBOP} in form of several parametric laws. A linear or quadratic limb-darkening law introduces another one or two parameters. Initial guesses of known quantities have been obtained from the respective discovery papers for each system.

When using \texttt{JKTEBOP\_V25} in its sim\-plest form \mbox{(TASK3)} the parameter uncertainties are obtained from the best-fit covariance matrix. Therefore they  
are formal (and probably too optimistic) errors. More proper parameter uncertainties are obtained through the implementation of bootstrapping, Monte-Carlo and residual-permutation (RP) algorithms. To obtain robust parameter uncertainties for light-curves with correlated red noise the RP algorithm is most suitable and available via TASK 9.

\section{Results - 1}
Our first few test observations with the telescope in focus aimed at recording transit light-curves of XO5 \citep{XO5DiscPap}, XO3 \citep{XO3DiscPap} and XO4 \citep{XO4DiscPap}. The resulting light-curves are shown in Fig.~\ref{XO5_LC}, Fig.~\ref{XO3_LC} and Fig.~\ref{XO4_LC}, respectively. In all three cases the transit is clearly visible. The photometric point-to-point scatter is no less than $\sim$ 3 mmag with maximum exposure times ranging from 105 seconds for XO5 (V = 12.1 mag) to 12 seconds for XO3 (V = 9.9 mag). Short exposure times result in a high sampling cadence. The residuals in each plot all show some degree of systematic effects indicating that correlated (red) noise is the most dominant error source. In all cases the photometric precision is too low for a sufficient characterisation of the transit. The derived planet parameters have a large uncertainty. Nevertheless we modelled the lightcurves in order to obtain an estimate of the mid-transit time. Limb-darkening treatment is not meaningful since no such effect is readily apparent in the time-series photometry. Errorbars per observation has been omitted in the figures to highlight some details concerning our observing strategy, sky condition and telescope tracking properties. 

In Fig.~\ref{XO5_LC} we also plot the used exposure time during the observing window. Depending on sky condition we varied the exposure between 75 seconds and 110 seconds. After BJD 2,455,968.15 the scatter systematically shifts upwards in the differential magnitude plot. This is coinciding with a period of time where the exposure time was chosen to be minimal demonstrating that decrease in exposure time deteriorates the photometric precision. This finding motivated us to stick to a constant exposure time during the course of observation resulting in a homogeneous data set.

In Fig.~\ref{XO3_LC} we chose to plot the air-mass during the XO3 transit. We demonstrate that observing through a relatively high air-mass ($>$1.45) results in a significant increase in photometric scatter most likely due to a larger scintillation noise. To obtain high-precision photometry of a transiting planet one should aim to observe through as low air-mass as possible. This requirement would exclude targets located on more southern latitudes. 

A last characteristic is concerned with the pointing ability of the telescope itself. The cross-correlation algorithm as implemented in \texttt{DEFOT} for image registration records the number of pixel with which each program image was shifted in order to match the reference image. In Fig.~\ref{XO4_LC} we plot the telescopic x- and y-drift measured in pixels during the XO4 transit observation. The x-drift corresponds to $\pm$ righ-ascension while y-drift is $\pm$ declination. In both directions the maximum drift is around 10 pixels (10.5 arcseconds). While the y-drift appears to be of continous nature the x-drift occurs in jumps. Pointing imperfections can either be due to slag in the motor drives/friction-disk, inaccurate encoder and/or software control. The ability to keep the target on the same pixels during the transit observations contributes to minimising flatfielding errors. As an example the recently commissioned K2 mission (Kepler space telescope extended mission) will have its photometric precision decreased significantly (compared to the original Kepler mission) due to decreased pointing accuracy.

\begin{table}[t]
\begin{center}
\centering
\caption{Theoretical limb-darkening coeffiencients for a quadratic LD law valid for a Cousin R-filter. See text for details.}
\begin{tabular}{cccc} 
\hline
Host star & linear $(u_a)$ & quadratic $(v_a)$ \\
\hline
\hline
HATP25    & 0.3879 & 0.2906 \\
HATP09    & 0.5588 & ------ \\
HATP22    & 0.4285 & 0.2646 \\
HATP12    & 0.5669 & 0.1687 \\
\hline 
\end{tabular}
\label{LDCoefficients}
\end{center}
\end{table}

\section{Results - 2}

In order to obtain high-precision photometric measurements of a transiting planet we then carried out tests with long exposure times per observation frame. A significant amount of telescope defocus would be required for the more brighter targets to ensure ADU counts to be in the linear regime of the CCD. We obtained four light-curves of four different systems on four different nights. Three of the recorded light-curves makes use of the new ST-16083 CCD camera. In Fig.~\ref{HATP25_LC} to Fig.~\ref{HATP12_LC} we show the light-curves of HATP25 \citep{HATP25DiscPap}, HATP9 \citep{HATP09DiscPap}, HATP22 \citep{HATP22DiscPap}, HATP12 \citep{HATP12DiscPap}. Light-curves of HATP12 were previously recorded several times using the Korean 1m optical telescope at Mt. Lemmon Optical Astronomy Observatory (LOAO), Arizona, USA \cite{Lee2012}. All light-curves were observed through a Cousins R-filter. The sampling frequency is now due to a longer exposure time. Exposure times of over 3 minutes have been used on average. The effect is immanent and we see a significant increase in photometric precision compared to our first test observations (see previous section). Using a fixed exposure time of 260s we achieved a RMS scatter of $\sim$ 1.85 mmag for the faintest host star (HATP12, V = 12.8 mag). A remarkable RMS scatter of 0.70 mmag was obtained for the brightest target HATP22 (V = 9.8 mag).

\begin{figure}[!t]
\centering
\epsfxsize=8.5cm 
\epsfbox{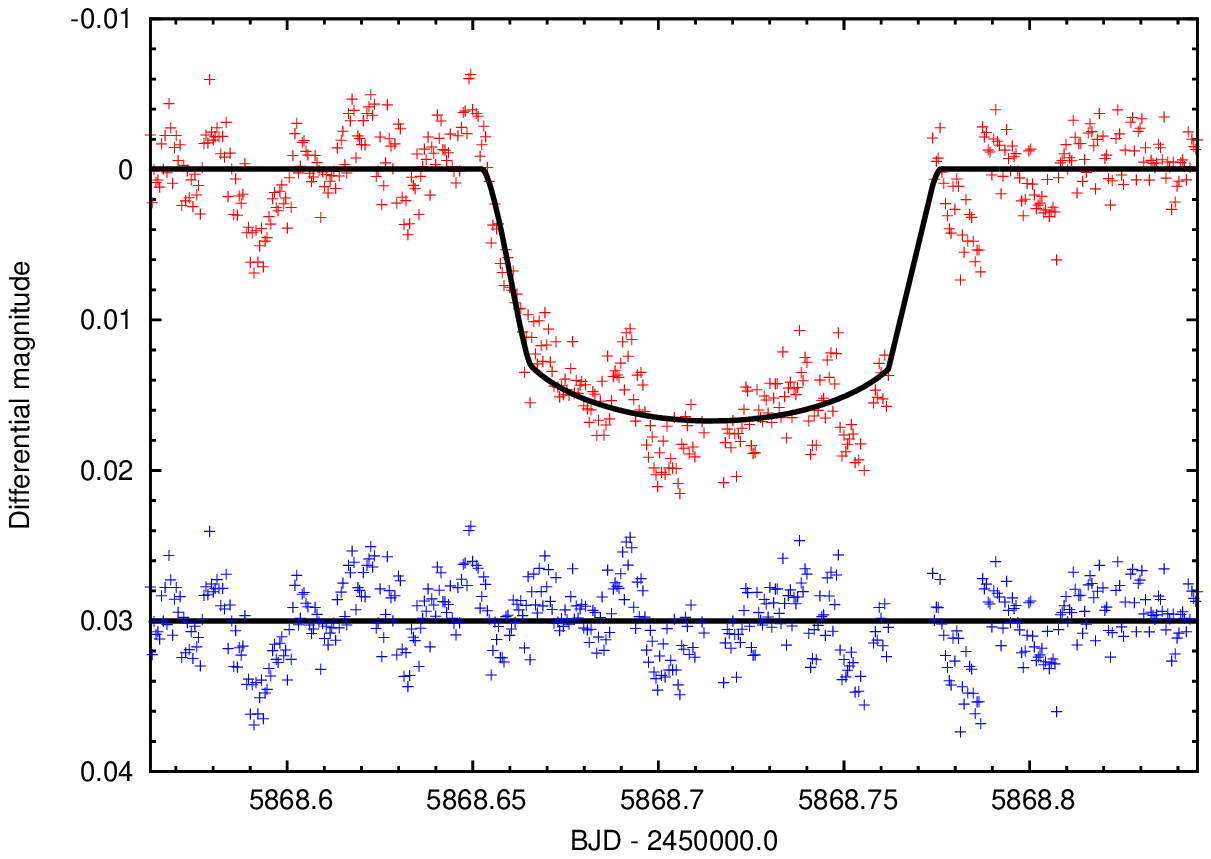}
\epsfxsize=8.5cm 
\epsfbox{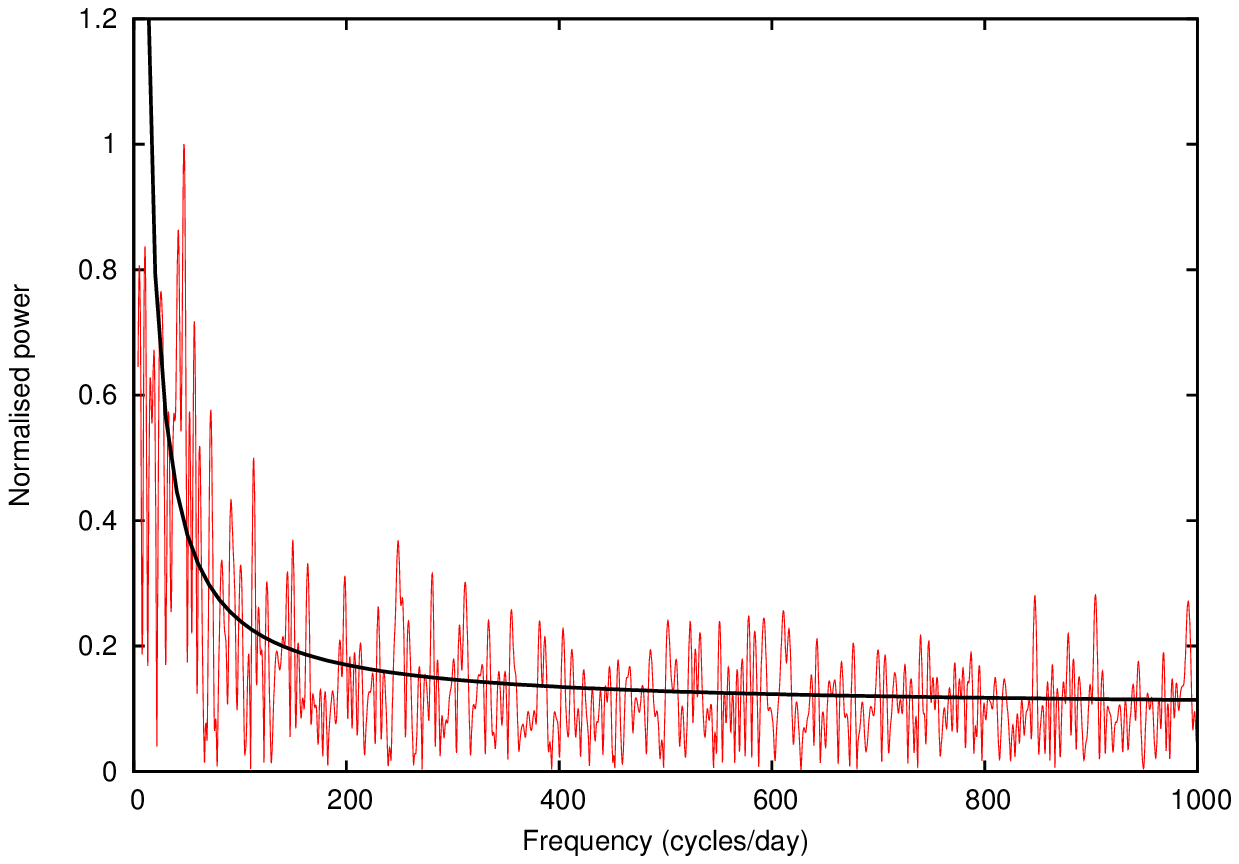}
\caption{Top panel: Light-curve of HATP16 as observed on the night of 2011/11/03 with the 1m reflector at Mt. Lemmon Optical Astronomy Observatory (LOAO). The RMS scatter around the best-fit model is 2.41 mmag. Lower panel: Power spectrum (normalised to the highest peak) of the residuals between observations and best-fit model. Clearly a 1/frequency dependency is evident.}
\label{HATP16_LOAO_LC}
\end{figure}

\subsection{Treatment of limb-darkening}

Quadratic limb-darkening coefficients for Cousin R-filter were obtained using the \texttt{JKTLD}\footnote{http://www.astro.keele.ac.uk/jkt/codes/jktld.html} code which outputs theoretically calculated limb-darkening strengths for a variety of parametric laws. The coefficients are calculated from bilinear interpolation (in effective temperature and surface gravity) from published tables calculated from stellar model atmospheres. In this work we obtained the coefficients for the quadratic limb-darkening law from tables published in \cite{Claret2000}. In all cases we assumed the metallicity of the host-star to be zero and the atmospheric micro-variability parameter were set to unity. Using nominal published values for $T_{eff}, \log g$, [Fe/H] and $v_{mic}$ does not allow one to determine suitable coefficients.

\subsection{Treatment of parameter uncertainties}

\begin{figure*}[!t]
\vbox{
\centerline{
\epsfxsize=8.5cm
\epsfbox{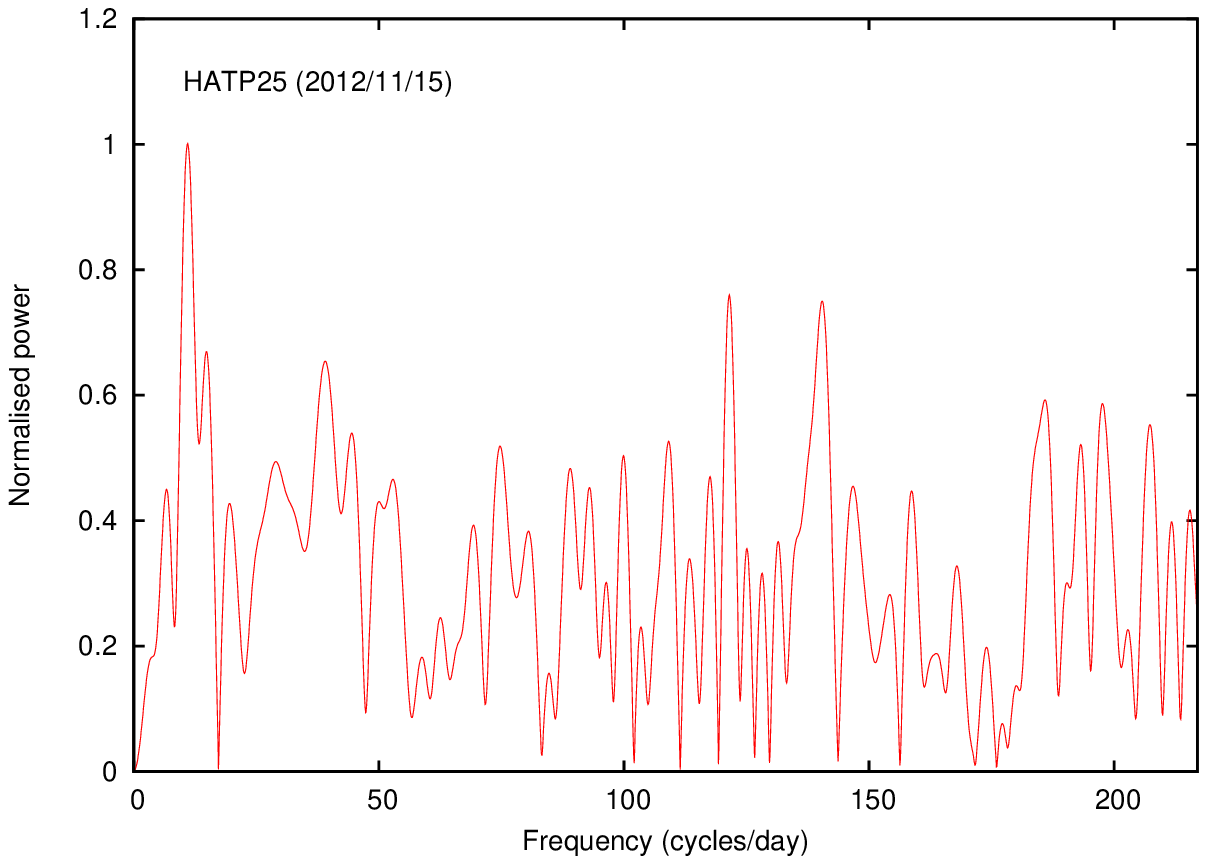}
\epsfxsize=8.5cm
\epsfbox{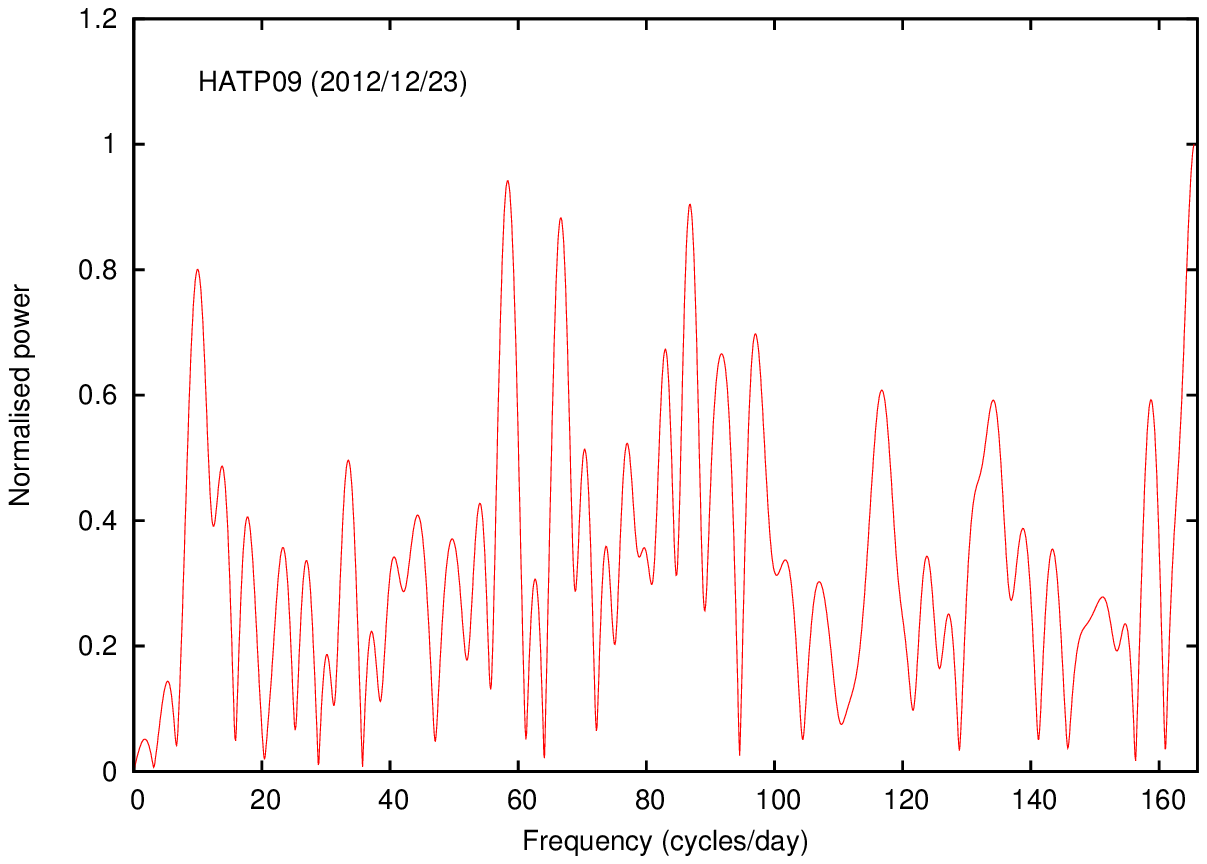}
}
}
\vbox{
\centerline{
\epsfxsize=8.5cm
\epsfbox{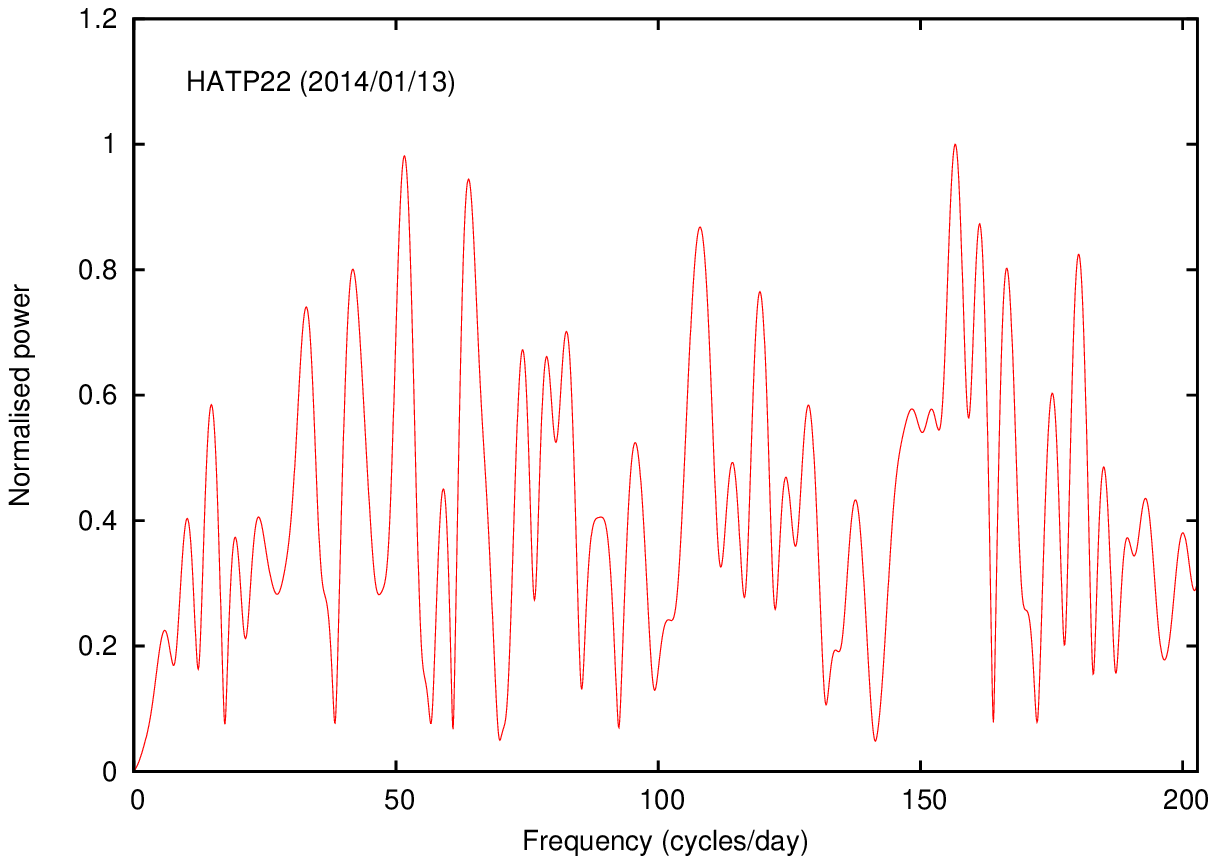}
\epsfxsize=8.5cm
\epsfbox{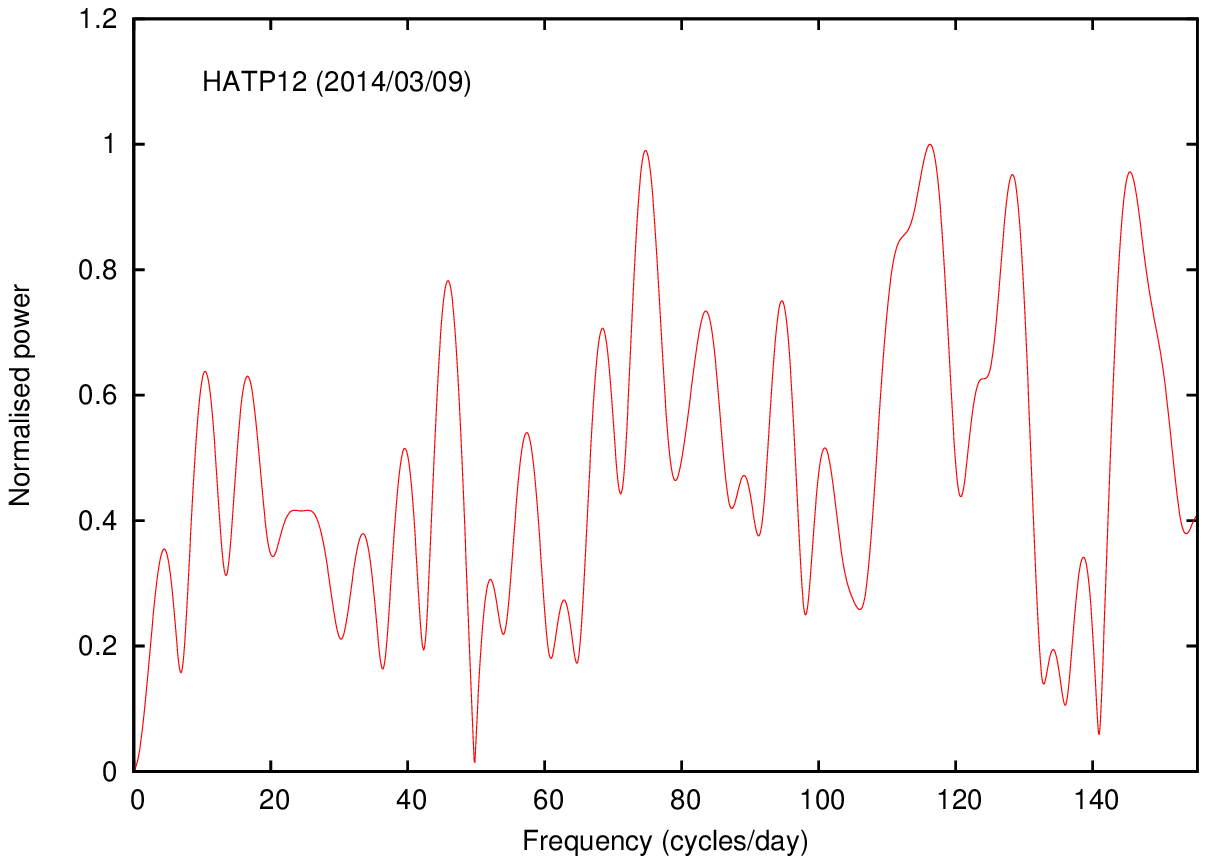}
}
}
\caption{Power spectral density plots for the time-series data of four transit residuals. Frequencies have been plotted up to the maximum Nyquist frequency. No obvious 1/frequency dependency is visible except (with some good will) for the transit of HATP25.}
\label{PowerSpectra}
\end{figure*}

\texttt{JKTEBOP} allows parameter uncertainties to be obtained from two different methods depending on the noise character in the data set. If the observational noise is time-correlated (red noise or covariant noise, see eventually \cite{Gillon2009}) then TASK9 is recommended and makes use of the residual-permutation (a variant of prayer-bead) algorithm. If the noise is Gaussian or white noise then a standard Monte-Carlo algorithm is implemented via TASK8. The former method requires the computation of $n$ shifted fits where $n$ is the number of data points. The latter method requires a larger number of simulations and we used 10,000 experiments to obtained parameter uncertainties from TASK8. The remaining question is how to judge whether a given data set contains red or white noise. One method is to calculate the power spectral density of the photometric time-series or its residuals from a best-fit solution. If the observations have a high content of time-correlated (red) noise then the power spectrum of the timer-series will exhibit a 1/frequency dependence: most power at low frequencies. To test this behaviour we calculated the power spectrum of a single HATP16 transit observed with the 1m reflector at Mt. Lemmon Optical Astronomy Observatory. Figure \ref{HATP16_LOAO_LC} shows the observed transit along with a best-fit solution. Systematic trends in the residuals are clearly visible during the observing window. The lower panel shows the power spectrum computed from a Lomb-Scargle algorithm as implemented in the \texttt{PERIOD04}\footnote{https://www.univie.ac.at/tops/Period04/} software package \citep{LenzBreger2005} with a clear 1/frequency trend pointing towards a significant amount of correlated noise (also visible in the residual plot in Fig.~\ref{HATP16_LOAO_LC}). We have calculated power spectra for the four planetary transits obtained with the CbNUOJ telescope and display the results in Figure \ref{PowerSpectra}. In all cases (with the HATP25 transit as a possible exception with a marginal content of red noise) the power spectrum is flat (certainly no 1/frequency dependency) and hence no time-correlated red noise is present in these data sets. We have therefore decided to use JKTEBOPs TASK8 for all transit observations to estimate parameter uncertainties. Since HATP25 could have red noise components we have also estimated parameter uncertainties using TASK9.

For HATP25 we first fitted for the two limb-darkening coefficients separately. We find that the data does not allow a simulteaneous fitting for the coefficients and found that the sum of final coefficients is larger than unity (implying an unphysical brightening at the hoststar limb). We therefore decided to fix those parameters at their theoretical values as shown in Table \ref{LDCoefficients}. The final best-fit parameters for the HATP25 transit is shown in Table \ref{BestFitParams}. Since we estimated parameter uncertainties using TASK8 and TASK9 we can now judge that some amount of time-correlated red noise is present for that dataset. The errors obtained from TASK9 are consistently larger than for TASK8.

For HATP09 we also fitted for the two limb-darkening coefficients separately, but encountered the same problem as for HATP25 (i.e unphysical limb-darkening coefficients). We then set the two linear and quadratic terms to their tabulated values as obtained from \texttt{JKTLD} and fixed those parameters. In this case we encountered that the inclination was kept artificially at 89.96 degrees to ensure numerical stability of the minimisation algorithm. However, the best-fit inclination in the discovery paper \cite{HATP09DiscPap} found an inclination of 86.5 degrees. We therefore decided to model the light-curve with a linear limb-darkening law using 0.5588 for the linear term as obtained from \cite{Claret2000}. The resulting best-fit parameters along with uncertainties are listed in Table \ref{BestFitParams}.

For HATP22 we judged that the amount of correlated red noise is minimal. We therefore used TASK8 (Monte-Carlo) to find parameter uncertainties based on 5000 simulations. The photometric precision is of high quality allowing us to fit for the two limb-darkening coefficients simultaneously without producing unphysical results. We also carried out fits where one of the parameters were held fixed. All three cases were consistent within the quoted uncertainties. The timing uncertainty for this light-curve is the smallest with around $\pm 17$ seconds. A close inspection of the HATP22 light-curve shows some systematic brightening at around BJD 2,456,671.215 and could be due to a star spot \citep{Nutzman2011, Oshagh2013}. However, this feature is also present in all the adopted comparison stars and hence is attributable to a short-term atmospheric change likely due to thin clouds.

For HATP12 we again inferred parameter uncertainties using TASK9. Also in this case we judged the errors to be dominated by correlated noise - especially prior to planetary ingress. We used a quadratic limb-darkening law with initial guess obtained from \cite{Claret2000} and listed in Table \ref{LDCoefficients}. Both parameters were kept freely adjustable during the initial fitting process and we did not encounter unphysical results. The final best-fit parameters along with their uncertainties are shown in Table \ref{BestFitParams}.

\begin{table*}[t]
\begin{center}
\centering
\caption{Final best-fit parameters for light-curves of HATP25, HATP09, HATP22 and HATP12. The mid-transit time $(T_0)$ is offset by BJD 2,456,000. The uncertainties have been obtained from TASK8 (Monte-Carlo) and TASK9 (Residual-Permutation) while setting the adjustable parameter integers to initially 1.}
\begin{tabular}{cccccc} 
\hline
Parameter       & HATP25                                        & HATP25                                    & HATP09                                     & HATP22                                   & HATP12                                    \\
\hline
\hline
$r_{A}+r_{b}$	& $0.136^{+0.009}_{-0.010}$           & $0.14^{+0.010}_{-0.020}$            & $0.126^{+0.009}_{-0.010}$          & $0.34^{+0.02}_{-0.02}$             & $0.28^{+0.02}_{-0.01}$              \\
$k$		     & $0.137^{+0.003}_{-0.003}$           & $0.137^{+0.005}_{-0.004}$          & $0.106^{+0.002}_{-0.002}$          & $0.106^{+0.002}_{-0.002}$        & $0.141^{+0.006}_{-0.006}$         \\
$i (^{o})$          & $85.3^{+0.8}_{-0.7}$                   & $85.0^{+1}_{-0.8}$                      & $87^{+1}_{-0.9}$                        & $84^{+5}_{-3}$                          & $88^{+2}_{-5}$                           \\  
$T_{0}$ (BJD)  & $247.1288^{+0.0007}_{-0.0007}$ & $247.129^{+0.001}_{-0.0007}$     & $285.1671^{+0.0005}_{-0.0005}$ & $671.2504^{+0.0002}_{-0.0002}$& $726.1710^{+0.0004}_{-0.0004}$ \\
$u_a$              & $0.3879$ (fixed)                          & $0.3879$ (fixed)                          & $0.5588$ (fixed)                         & $0.6^{+0.2}_{-0.2}$                    & $0.7^{+0.3}_{-0.3}$                    \\
$v_a$              & $0.2906$ (fixed)                          & $0.2906$ (fixed)                          & $-$                                            & $-0.1^{+0.4}_{-0.4}$                   &  $0.0^{+0.7}_{-0.8}$                   \\
$r_A$              & $0.119^{+0.008}_{-0.009}$          & $0.121^{+0.009}_{-0.014}$           & $0.114^{+0.008}_{-0.009}$        & $0.31^{+0.01}_{-0.01}$                &  $0.24^{+0.02}_{-0.009}$            \\
$r_b$              & $0.016^{+0.001}_{-0.001}$          & $0.016^{+0.002}_{-0.002}$           & $0.012^{+0.001}_{-0.001}$         & $0.033^{+0.002}_{-0.002}$          &  $0.034^{+0.004}_{-0.0007}$       \\
$\sigma$ (mmag) & $2.30$                                   & $2.30$                                        & $1.68$                                       & $0.70$                                        & $1.85$                                       \\
$\chi_{r}^2$         & $1.01$                                   & $1.01$                                         & $1.42$                                      & $1.21$                                        & $0.91$                                       \\
JKTEBOP & & & & &\\
uncertainty method         & TASK8                          & TASK9                                            & TASK8                                        & TASK8                                          & TASK8                                          \\
\hline 
\end{tabular}
\label{BestFitParams}
\end{center}
\end{table*}

\section{Conclusion}

We have presented first results from follow-up observations of several transiting planets using the CbNUOJ 0.6m telescope. We have recorded light-curves using infocus as well as defocus telescope settings. We have qualitatively described various photometric noise sources and their effects when operating the telescope in a defocus mode. In particular we have outlined the advantage of defocus observations and demonstrated to achieve a precision of RMS $\sim$ 1 milli-magnitude over a time-scale of several hours. We believe the precision can be explained by three factors: 1) long integration time (enabling the collection of many photons), 2) controlling systematic noise source due to good pointing precision by stable telescope tracking (keeping the target star on roughly the same pixels) and 3) beating of scintillation noise (as a side-effect from long exposure times) which otherwise would be significant for a 0.6m telescope using exposure times of a few tenth of seconds when operating in in-focus mode. We demonstrated that the telescope is well suited for follow-up observations of suspected transit signals from discovery surveys such as Qatar, HAT, WASP and similar on-going projects. However, target stars with possible planets have to be bright with a limiting magnitude of around V = 13 - 14 (depending also on the transit depth). For bright host stars with V $\sim$ 10 a remarkable photometric precision of RMS $\sim$ 0.70 milli-magnitude was demonstrated for HATP22. This precision is high enough to determine physical properties of the planet with errors of a few percent. Unfortunately we recorded only one complete transit of HATP22. Additional datasets would be necessary for a complete analysis also helping to reduce modelling errors. Furthermore, tracking of star spots visible during a transit event can be used to measure stellar rotation of the host star and would be detectable with a sub-milli-magnitude precision. Mid-transit times were measured to $\sim$ 17 seconds for HATP22 to $\sim$ 86 seconds for HATP25 enabling some or limited opportunity for transit-timing variation follow-up studies for the detection of additional bodies in the system. In the future we plan to carry out follow-up observations of bright host-stars known to harbour a transiting planet with the aim to obtain $\sim$ 1 milli-magnitude photometric precision.

\acknowledgments{The data acquisition and analysis was partially supported by Basic
Science Research Program through the National Research Foundation of Korea (NRF) funded by the Ministry of Education, Science and Technology (2011-0014954). 
C-H K was partially supported by the Korea Research Foundation (KRF) grant funded by the Korea Government (NRF-2012R1A1A4A01012467).}


\end{document}